\documentclass[useAMS,usenatbib]{mn2e}
\usepackage{epsfig,graphicx,lscape}
\title[WASP ExtraSolar Planet Candidates]{The SuperWASP wide-field exoplanetary transit survey: Candidates from Fields 23hr $<$ RA $<$ 03hr} 
\author[D.J. Christian et al.]{D.J. Christian$^{1}$,
D.L. Pollacco$^{1}$,
I. Skillen$^{2}$,
R.A. Street$^{1}$,
F.P. Keenan$^{1}$,
\newauthor
W.I. Clarkson$^{3}$,
A. Collier Cameron$^{4}$,
S.R. Kane$^{4}$,
T.A. Lister$^{4,6}$,
R.G. West$^{5}$,
\newauthor
B. Enoch$^{3}$, 
A. Evans$^{6}$,
A. Fitzsimmons$^{1}$,
C.A. Haswell$^{3}$,
C. Hellier$^{6}$,
\newauthor
S.T. Hodgkin$^{7}$, 
K. Horne$^{4}$,
J. Irwin$^{7}$,
A.J. Norton$^{3}$,
\newauthor
J. Osborne$^{5}$,
R. Ryans$^{1}$,
P.J. Wheatley$^{8}$, 
and 
D.M. Wilson$^{6}$  \\
$^{1}$Astrophysics and Planetary Science Research Division,\\ Department of Physics and Astronomy, Queen's University Belfast, Belfast, BT7 1NN, UK\\ 
$^{2}$Isaac Newton Group of Telescopes, Apartado de correos 321,
E-38700 Santa Cruz de la Palma, Tenerife, Spain\\ 
$^{3}$Department of Physics \& Astronomy, The Open University, Milton Keynes, MK7 6AA, UK  \\
$^{4}$School of Physics \& Astronomy, University of St. Andrews, North Haugh, St. Andrews, Fife, KY16 9SS, UK  \\
$^{5}$Department of Physics \& Astronomy, University of Leicester, Leicester, LE1 7RH, UK \\ 
$^{6}$Astrophysics Group, School of Chemistry \& Physics, Keele University, Staffordshire, ST5 5BG, UK  \\
$^{7}$Institute of Astronomy, University of Cambridge, Madingley Road, Cambridge, CB3 0HA, UK \\
$^{8}$ Department of Physics, University of Warwick, Coventry CV4 7AL, UK
}  

\date{Accepted 2006 August 3. Received 2006 August 3; in original form 2006 July 3}

\begin{document}

\maketitle

\label{firstpage}

\begin{abstract} 
Photometric transit surveys promise to complement the currently known
sample of extra-solar planets by providing additional information on the
planets and especially their radii. Here we present extra-solar planet 
(ESP) candidates from one such survey called, 
the Wide Angle Search for Planets (WASP) obtained
with the SuperWASP wide-field imaging system.  Observations
were taken with SuperWASP-North located in La Palma during the April to October 
2004 observing season.  The data cover fields between 23hr and 03hr
in RA at declinations above +12. This amounts to over $\approx$400,000 stars 
with V magnitudes 8 to 13.5. For the stars brighter than 12.5, we  
achieve better than 1 percent photometric precision. Here we present 
41 sources with low amplitude variability between $\approx$ 1 and 10 mmag,
from which we select 12 with periods between 1.2 and 4.4 days as
the most promising extrasolar planet candidates.
We discuss the properties of these ESP candidates, the expected fraction of transits 
recovered 
for our sample, and 
implications for the frequency and detection of hot-Jupiters.

\end{abstract}



\section{Introduction} 
The discovery of the large debris disk around $\beta$ Pic \citep{ST84}
and other young stellar sources with disks
\citep{J98, L97},
plus the subsequent discovery of numerous extra-solar planets (ESP) has led to a
resurgence in the study of planet and solar system formation. 
Discoveries of ESPs
have been dominated by radial velocity surveys \citep{MQ95, MB98, MB00, Ud00}. 
Currently, there are over 170 known ESPs orbiting around over 150 host stars. 
Photometric transit surveys have identified 
about 10 new planets, starting with: the confirmation of the first system,  
HD 209458b \citep{Hen00, Ch00}, 
followed by several discovered with the OGLE project 
and confirmed with radial velocity 
follow-up \citep{Kon03, Kon04, Po04, Bo05a, Kon05}, 
to the discovery of TrES~1 \citep{Al04}, 
the detection of the transits from the radial-velocity 
discovered HD~189733 \citep{Bo05b}, and 
the recent discovery and follow-up of XO-1b \citep{Mc06, W06}. 
These planets typically have periods less than 4 days,
orbital distance 
$\sim$0.05 AU, and Jupiter-like masses and radii, and hence
have been called the {\it hot-Jupiters}. Additionally, several very short 
period (P $<$ 1.5 d) systems have been discovered \citep{Kon03,Bo04} 
and termed {\it very-hot-Jupiters}. 

Radial velocity (RV) surveys measure the Doppler velocity of individual 
stars at many different times, which is a very time
consuming procedure. The photometric transit method can measure thousands of
stars simultaneously. 
In addition to monitoring a large number of stars, photometric transit
detections provide information on the planet's radius and inclination,
not readily available from RV surveys. 
It is currently not know what fraction of stars contain
planets. The fraction of planets that may be hot-Jupiters is also not
known.   
About 20\% of the currently known ESPs have periods less than 10 days
\citep{B06} and 
$\approx$10-15\% of solar-type stars surveyed have ESPs \citep{LG03}. There have been 
several attempts to better define what percentage of stars have planets. 
These range from 1-2\% \citep{Br03} to 20\% \citep{LG03}, 
but the latter also speculate this 
may be a lower limit. The large number of stars covered by photometric transit 
surveys should provide
better constraints on the fraction of stars with
hot-Jupiters, and may shed light on their formation mechanisms. 

Here we present results from one such photometric transit survey,
called WASP (the Wide Angle Search for Planets). WASP consists of 2 separate
telescopes, called SuperWASP North (SW-N) and SuperWASP South (SW-S).  The
SuperWASP telescopes were designed to cover a large field-of-view with better
than 1\% photometric accuracy down to 12th magnitude. Each SW telescope consists
of 8 separate cameras, each of which covers 61 sq degrees of sky, 
and the combined 
system surveys nearly 500 sq degrees. The results presented here are 
from the 2004
SW-N observing campaign taken with a 5 camera configuration. A description of
the SW instruments and observing strategy are presented in \S\ 2. In \S\
3 we present the data reduction and analysis techniques for the ESP search.
Candidates with secure low amplitude variability that may be ESPs are presented
in \S\ 4. In \S\ 5, we discuss possible false alarms, expected fraction of 
recovered transits, and indications for the
frequency of hot-Jupiters, their properties and observing follow-up strategies
to confirm the transits planetary nature. 
Lastly, in  \S\ 6 we summarize our findings.

\section{Instrumentation and Observations} 
\subsection{Instrumentation}
The SuperWASP (SW) telescopes were designed to cover a large area of sky and achieve 
photometric accuracy of a few millimags, and improve on   
the success of the prototype WASP0 instrument \citep{Kane04, Kane05}. 
SW~N in La Palma is contained in its own custom enclosure
with a hydraulically operated roll-away roof, and its own GPS and weather
station \citep{Ch05, P06}.  
The telescope mount is a rapid slew fork mount ($\approx$10$^{o}$/sec) from OMI, Inc. 
Commercially available  
components were used where available to keep costs down and decrease construction time. 
These components include: the telescope control system (TCS), a dedicated PC for each camera 
for data acquisition, and a PC attached to a tape (DLT) autoloader for data storage. 
The TCS monitors the weather, sets time for the FITS
data header from the GPS and can close the roof in the event of a weather
alert \citep{P06}. 
All major components are run with Linux OS.

To meet the science requirements of covering a large area of sky with high
photometric precision, a combination of Canon 200-mm f/1.8 lenses and Andor
e2v 2k$\times$2k back illuminated CCDs was chosen. The CCDs are passively cooled with a
Pelletier cooler and have a very short 4-sec readout time. This combination of
lens and camera gives a field of 7.8$^o$ $\times$ 7.8$^o$  ($\sim$61 sq
degrees). No additional filters were used and this ''wide-V'' set-up covers 
from 8--15 magnitude for a typical 30 second
exposure.  A $<$1\% photometric precision was obtained for magnitudes brighter than V=12.5.
Additional details of the SuperWASP project and instrumentation is presented
in \citet{P06}.

\subsection{Observational Strategy} 
SW-N was commissioned in November 2003, 
and inaugurated on 16 April 2004. 
Our initial observing strategy was tailored
toward searching for extra-solar planet transits.  This was to observe fields
with a large number of stars, but to avoid the Galactic plane where over-crowded
fields would increase the number of blended stars and make detection and data reduction 
difficult. Based on the Besancon
Galactic model \citep{RR03} a declination of +28 was chosen, with the telescope stepping
through RA in 1 hour increments centered on the current LST, but within the
$\pm$4.5 hr hour angle limit of the mount.  A maximum of 8 fields were observed
with a duration of $\approx$1 minute per field, including a 30
second exposure, 4 second read-out and the time for the telescope to slew and
settle at the new field.  Such observations provide well sampled light curves
with a cadence of better than 8 minutes between subsequent measurements.  
This observing strategy provides
a nightly baseline of over 6 hours of coverage for the 6 fields centered near LST at midnight, and
over 4 hrs of coverage on $\approx$10 fields per night.  A typical field contains
$\sim$20,000 stars per camera at magnitudes brighter than 13.  
Nightly calibrations, such as bias levels, darks, and flat fields to 
remove the pixel-to-pixel variations and vignetting were obtained
using automated scripts. Biases and flat fields were obtained at
the beginning and end of each night, and flat fields 
were taken during twilight with the telescope pointed at zenith and the drive off. 
Exposure times were initially based on the data in \citet{TG93} and
adjusted from the counts in the first image.

The $\approx$ 150 nights from the 2004 observing season covered several
dozen separate fields and resulted in the detection of slightly 
more than 6.7 million stars. For the extra-solar planet search these observations 
were separated into 6 equal sections of the sky, each containing about 
1 million stars. This work presents the stars from 23hr $<$ RA $<$ 03hr,
and contains nearly 900,000 stars observed for at least 25 nights and 
$\approx$400 image frames. The statistics for the observed fields are presented
in Table~1, and the distribution of all stars in our sample as a function
of SW-N V magnitude is shown in Figure~1. Approximately 400,000 stars are
brighter than 13.5 magnitude.

\begin{table*} 
 \centering
 \begin{minipage}{65mm}
\caption{Field Statistics} 
\begin{tabular}{lccc} 
\hline 
\hline 
Field &  Nights   & Images  & Stars   \\
SWASP~J &   \\\hline 
2317+2326 & 115 & 4882 & 38993  \\
2343+3126 & 112 & 4569 & 51694  \\
2344+2427 & 92  & 3406 & 38332  \\
2344+3944 & 110 & 4539 & 59202  \\
0016+3126 & 111 & 4525 & 48448  \\
0017+2326 & 110 & 4501 & 36676  \\
0043+3126 & 29 & 406 & 38458  \\
0044+2127 & 61 & 2046 & 30156  \\
0044+2427 & 29 & 406 & 34267  \\
0044+2826 & 80 & 3088 & 40578  \\
0045+3644 & 80 & 3061 & 46856  \\
0115+2826 & 79 & 3063 & 33249  \\
0116+2027 & 78 & 3051 & 27183  \\
0116+3126 & 28 & 372 & 34276  \\
0117+2326 & 27 & 371 & 28312  \\
0143+3126 & 72 & 2591 & 36623  \\
0144+2427 & 52 & 1576 & 28059  \\
0144+3944 & 71 & 2570 & 56242  \\
0216+3126 & 71 & 2560 & 40368  \\
0217+2326 & 70 & 2559 & 30535  \\
0243+3126 & 61 & 1883 & 40191  \\
0244+2427 & 42 & 1017 & 28958  \\
0244+3944 & 60 & 1879 & 78643  \\
\hline
\end{tabular}
\end{minipage}
\end{table*}

\begin{figure}
    \epsfig{file=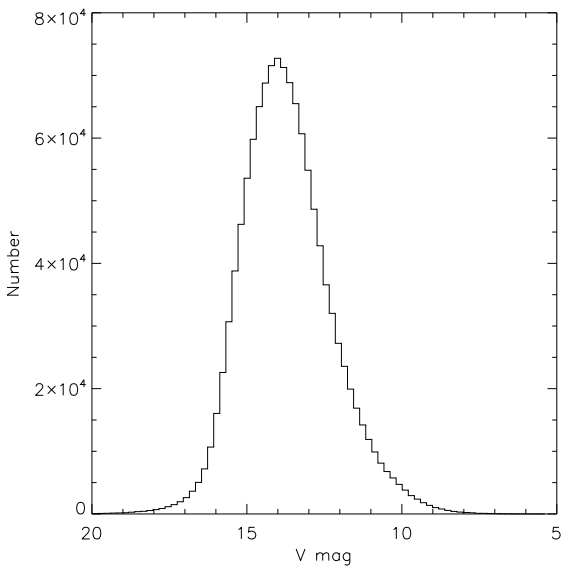, width=8cm} 
\caption{Distribution of stars in our sample (23 $<$ RA $<$ 03 hr) as a
function of V magnitude. 
\label{fig1}} 
\end{figure}

\section{Analysis} 
\subsection{Pipeline} 
We have constructed a custom data reduction pipeline, with a goal of
obtaining millimag photometric precision for stars with V $\leq$ 12.  
The pipeline (only briefly described here) uses custom FORTRAN programs 
combined with shell scripts and several STARLINK packages, and creates 
master biases, darks, and flat fields for each night of observations.  
Flat field calibration frames from individual nights are combined into a master flat using an exponential
weighting with the contribution of flats older than 14 days diminishing
\citep{CC06}.  Each science exposure is bias subtracted, dark
corrected, flat fielded, and the astrometric solution is computed using
reference stars from the USNO-B1.0 and Tycho catalogs.  

Aperture photometry is performed on the final calibrated images
with a custom-built package tailored
to deal with the ultra-wide fields. Bad pixel masks are applied to each frame,
and a blending index is assigned for every object detected.  
The blend index is defined as the ratios of the fluxes in various sized apertures
for an individual star \citep{P06}. We define two blend indices based on
the flux ratios: $r_1 = (f_3 - f_1)/f_1$ and $r_2 = (f_3
-f_2)/f_2$, where $f_1$, $f_2$, $f_3$ are the fluxes measured in 
apertures of 2.5, 3.5, and 4.5 pixels, respectively. 
Removing night-to-night variations in air-mass and sky conditions
was difficult, but the aperture photometry does achieve slightly
better than the required 1\% precision at V=12.5.
For the transit searching, additional systematic errors 
are removed with the {\em SysREM} algorithm \citep{Tam05}.
RMS precision as a function of
SW $V$ magnitude are shown in \citet{CC06} and \citet{P06}.   
Fluxes are computed in the 3 mentioned above apertures and output, 
along with other source attributes to a FITS binary table. 
The final pipeline data products include the calibrated images, 
catalog information, and time-tagged photometric data for each star. 
Trends (such as extinction) are removed from the data, and 
the binary table is ingested by the archive. 
A list of all cataloged targets is compiled from the
ingested FITS tables and compared with the WASP catalogue.
New objects are given IAU-compatible names, and all objects
are assigned to a 5$^o\times5^o$ sky tiles based on their coordinates.
Photometric data are split from the input files and stored
as intermediate per-sky-tiles. Photometric points within each
sky-tile are re-ordered to ensure that they are in consecutive rows
for a given star. These files are registered within the archive
data-base management system (DBMS) along with their names, location, and various meta-data.
The archive can then be interrogated and the photometric data
extracted as a single, coherent per-object light curve for
longer term temporal analysis. The object catalog is expected
to increase by $\geq 3\times10^6$ objects per year.
This public archive is hosted at Leicester within
LEDAS (Leicester Database and Archive Service).

\subsection{Transit Searching} 
The large number of stars makes visual inspection of every light curve
unfeasible. For this reason the transit searching programs provide rank ordered
lists of the best light curves based on derived transit depth, signal-to-noise,
and for periods less than 5 days. Another important and effective parameter in culling
the list of candidates was the signal-to-red noise, (S$_{red}$). S$_{red}$ is the
ratio of the best fit transit depth to RMS scatter and is described in detail
in \citep{CC06}.  
Our  refined list of candidates can then be
inspected visually for further analysis. Light curves that
show unusual extinction variations or other systematic problems can
then be eliminated.
Once the spurious candidates are culled from the master list further 
analysis of the light curves is undertaken.

There are several well used transit detection algorithms 
\citep{T03, Mo05}.
These include the box-shaped searches, such as, 
the Box-fitting Least-Squares (BLS) algorithm \citep{KZM02}, 
matched filter algorithms \citep{Kay98}, and Bayesian techniques 
\citep{DDP01, AF02}.
We have applied an improved
BLS algorithm to the 2004 data to search for 
transits. This method is described in detail in a companion paper by 
\citet{CC06}.  BLS allows fast and efficient searching
for transits, while keeping the false alarm rate low. 
Additionally, \citet{Mo05} suggest that the BLS method was preferable 
in a comparison of several commonly used algorithms.  
An extensive comparison of the leading transit detection algorithms as applied to SuperWASP
data is currently underway \citep{E06}.   

Several additional techniques were adopted to the  
transit finding software to keep the false alarm rate low. These include a test
for ellipsoidal variations in the light curve, such as would be produced
by close stellar eclipsing binary systems. The amplitude of the 
ellipsoidal variations,  (S/N)$_{ellip}$, is also given for each system. 
Similar methods applied to OGLE candidates
were successful in distinguishing stellar sources from 
true ESP candidates \citep{SP03}. 
Additionally, true ESP transit light curves
should be flat at the phase 180$\degr$ away from the transit, 
at phase 0.5, the 'anti-transit'. Hence
tests for the variability at phase 0.5 \citep{Bu06} were also added.
We present our results for this set of SW data in the next section. 
From these and further scrutiny to distinguish low amplitude stellar systems
from true ESPs we cull our list of extra-solar planet candidates. 

\section{Results}

Systems that have good signal-to-noise and transit depths 
less than 10\% are presented in Table~2. Here we have excluded
systems showing secondary eclipses in their light curves and
stars with any brighter neighboring stars within a 48$\arcsec$ radius. 
We will apply strict criteria to the sources listed in the table to separate the
stellar systems from the possible ESPs. 
Systems likely to be low stellar mass binaries 
are interesting in their own right and if confirmed would add valuable 
information to the Mass-Radius relations for low mass stars, 
and are thus included here. 
However, our current analysis does not allow us to give an accurate estimate of
the binary fraction of the SuperWASP fields. 
Our transit searching methods were optimized for 
finding transit-like events, and  
many types of binaries (those with large changes in magnitude ($>$10\%) or large
ellipsoidal amplitudes) are not reported. Thus, any estimate would
only provide a lower limit to the true binary fraction, and this 
investigations is left for future publications.

\subsection{ESP Candidate Selection} 
In the present work we are
interested in the best possible ESP candidates and require these
to have: (i) high signal-to-noise light curves (S$_{red} >$ 7), 
(ii) a reasonable transit depth for spectral type,  
(iii) a reasonable transit duration, with $\eta < $ 1.3, where
$\eta$ is the ratio of the observed transit duration
to that expected based on the stellar and planetary radii \citep{TS05},
(iv) 
a flat bottomed transit shape, (v) to have been observed for at least 3 
transits, and (vi) an ellipsoidal amplitude less than 5.  Later type main-sequence
dwarf stars will have deeper eclipses, and transits depths 
may be as large as 20\%. Most hot-Jupiters have periods less than
10 days, which thus limits the transit duration to $<$ 7 hrs \citep{Kane05}.
A grazing angle transiting ESP may show a 'V-shaped' transit, but in
most cases this shape would indicate a grazing incidence eclipsing stellar
system, and these types of transits are further scrutinized.

In order to estimate the size of the transiting planet, the size
of the host star must also be determined. Several colour-temperature relations
were used to constrain spectral types and stellar temperatures. 
The V$_{SW}$ and USNO B1 and R1 magnitude are available for all stars
in our sample, and Tycho B and V plus 2MASS J, H, and K 
were used when available. 
Relations between V$_{SW}$ - K and temperature were derived using a coarse
grid of V and K colours from a selection of 30,000 stars from 
\citet{Am06}. 
Subsequently, the stellar radii were determined from the derived stellar 
temperatures using J-H colors and the relation from \citet{Gr92}.
This relation was further optimized with a polynomial fit for the FGKM 
temperature range. No reddening correction was included in V$_{SW}$ - K.
In general, with SuperWASP observations are away from the Galactic plane,
the reddening correction should be small, but would serve to increase
V-K and artificially lower the derived stellar radii.
In Figure 2., we show a histogram of stellar radii for a sample of $\approx$2,000 
stars (ones having all of the appropriate colours) selected from all fields and returned from the 
transit search code as having 'transit-like' variability.  The  
large peak in the figure at radii between 0.6 and 0.8 R$_{\sun}$ corresponds to the
late K and M0 stars from the faint end of the SuperWASP data.

 \begin{figure} 
    \epsfig{file=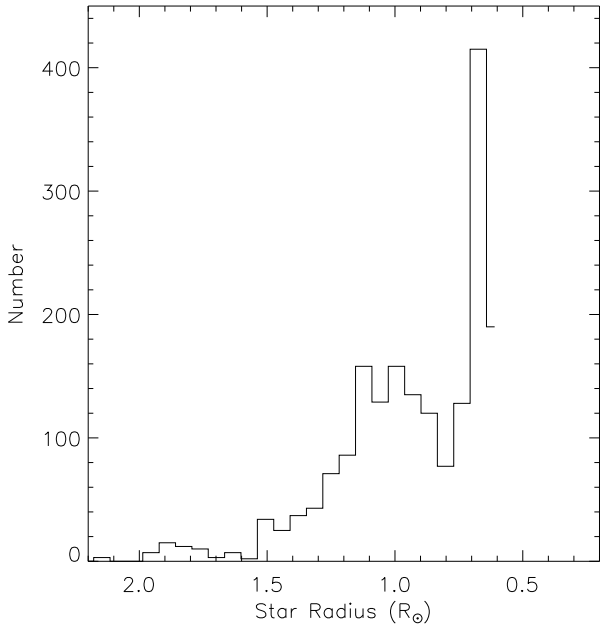, width=9cm} 
\caption{Histogram of stellar radii for a sample of stars 
from all fields returned from the transit search code (see text).
\label{fig2}} 
\end{figure}

We derived a planet's radius using  
the estimated stellar radius 
and a simple geometric transit model. 
The transit depth, $\Delta$F, is related to the 
ratio of the area of the planet to that of the star
with the equation:
 
\begin{equation}
\frac{\Delta F}{F} = (\frac{R_{p}}{R_{s}})^{2}
\end{equation}
 
where F is the baseline flux of the star,
and R$_p$ and R$_s$ are the radius of the planet and star, respectively. 
Planetary radii in Table~3 are given in units of Jupiter-radii using the
above relation from \citep{TS05}, and the table  
also includes $\eta$.

Additionally, we further refined the spectral type and host star's radii using
2MASS colours to distinguish between
dwarfs and giants. Similar transit depths for non-main sequence stars would
imply radii larger than that of Jupiter-sized planets. 
The 2MASS J-K color index is also presented in Table~3, and 
we compare the transit depths to infrared colors in Figure 3. J$-$K has
been shown to be a powerful tool in distinguishing dwarfs from giants
(Brown 2003). Values of J$-$K $>$ 0.7 are a strong indication that the
star is a giant, and hence the observed transit is caused by a body
much larger than a Jupiter-sized ESP. Or in some cases the giant may
be diluting the eclipse of a stellar binary, and also be an imposter.  
The 3 sources excluded as 'Giants' are over-plotted with filled circles.

\subsection{ESP Transit Candidates} 
The above criteria, (presented in \S\ 4.1) 
were rigorously applied to our
selection of the best low amplitude sources.
This leaves us with 12 good ESP candidates. 
Good ESP candidates 
are noted in bold face and commented as {\it ESPc} in Table~3. 
Over-plotted in Figure~3 are the theoretical curves for 0.5, 1.0 and 1.5 Jupiter radii planets. 
It can be seen 
many of these source's transits have reasonable planetary
radii between 0.5 and 1.5 R$_J$, and our best candidates are 
indicated by a filled $\star$ symbol. 
Other sources in the figure have large ellipsoidal variation or have
large derived planetary radii (R$_p$ $>$ 1.9 R$_J$),
and hence have been eliminated. 

 \begin{figure} 
    \epsfig{file=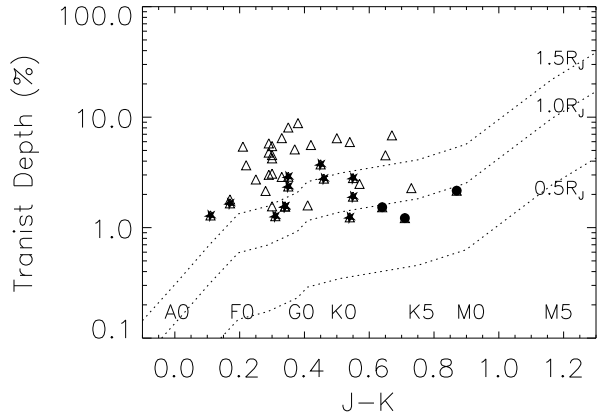, width=9cm} 
\caption{
Transit depth plotted as a function of colour indices, J$-$K
for our sample of sources (indicated as open triangles, $\bigtriangleup$), with the
most promising 12 candidates from Table~3 over-plotted as 
filled stars ($\star$). The 3 sources noted as 'Giants' are over-plotted
as filled circles. 
Over-plotted are curves showing the
expected transit depth for planets with radii of
0.5R$_J$, 1.0R$_J$, and 1.5R$_J$ (see text). 
\label{fig3}} 
\end{figure}

  We now discuss several of the sources that were not chosen as 
ESP candidates and those reasons, and we then discuss the properties of  
the best ESP candidates individually. 
Light curves and periodograms for these sources are shown in Figures 4--17.
Periodograms are included to show the robustness of the period searches and
are discussed for unusual cases. 

 Several sources presented in Table~2 had ellipsoidal amplitudes
great than 5, although several of these sources had reasonable
planetary radii.  We eliminated these as extra-solar planet candidates
and these are noted as {\it Ellip. large} in Table~3.  {\bf 1SWASP J000029.25+323300.9} and
{\bf 1SWASP J230808.34+333803.9} are two such sources 
and their folded light curves and 
periodograms are shown in Figures~4 and 5. 

 \begin{figure} 
    \epsfig{file=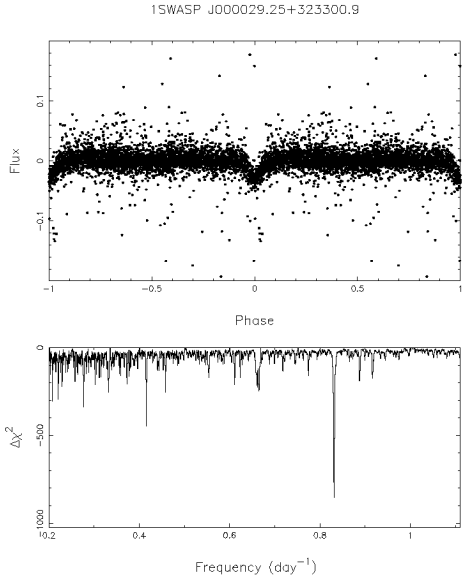, width=7cm} 
\caption{
We show sample timing data for sources that were disqualified as ESP candidates because
of large ellipsoidal variations. The 
{\it top} panel shows the light curve folded on the best fit period,
and {\it bottom} panel shows the periodogram for   
1SWASP J000029.25+323300.9.
\label{fig4}} 
\end{figure}
 \begin{figure} 
    \epsfig{file=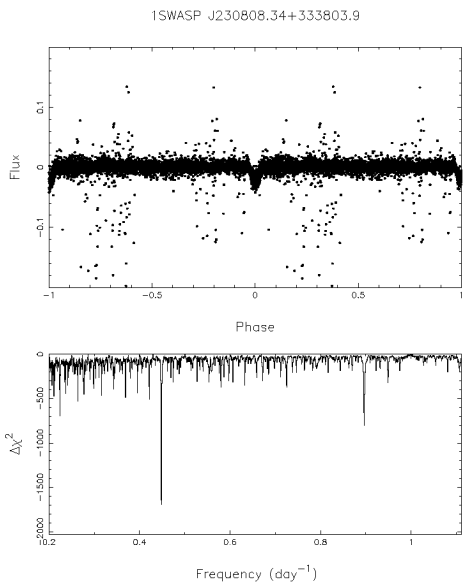, width=7cm} 
\caption{
Same as Figure~4 for 
1SWASP J230808.34+333803.9.
\label{fig5}} 
\end{figure}

Three sources that had derived planetary radii below 0.8 R$_J$ and also
passed most criteria, including a low ellipsoidal magnitude, but may very well
be giant and not dwarf stars. These 3 stars, {\bf 1SWASP~J002728.02+284649.2, 
1SWASP J013452.76+293626.5}, and {\bf 1SWASP~J231302.08+262724.3}
have J$-$K greater than $\approx$0.7. These stars also have very large values for $\eta$
and this may also indicate the object causing the transit is indeed stellar.

{\bf 1SWASP J002040.07+315923.7}  
A total of 9 transits were detected 
for this system, and the periodogram shows a well determined period of 2.55 days (Figure 6). 
We classify this as a G1 star using colour and temperature relations. 
This gives a planet radius of 1.2R$_J$. 
Although the transit is clearly visible in its light curve, there are 
several faint sources within the SW aperture and blending can not
be ruled out. 

{\bf 1SWASP J003039.21+205719.1}, {\bf 1SWASP J005225.90+203451.2},
and 
{\bf 1SWASP J010151.11+314254.7} were each observed in 2 different
cameras. Although their light curves and periodograms show some scatter, 
nearly identical transit depths and periods  
were derived for each star from both of their observations.
Their folded light curve and timing information are shown in Figure 7--9.
The planetary radii derived for these objects range from 1.2--1.4 R$_J$,
and the latter two stars have ellipsoidal amplitudes slightly greater than
3, and these may indeed be stellar systems. 
However, their ratios
of the observed to theoretical transit durations ($\eta$) are reasonable.

With a period of 4.4 days, {\bf 1SWASP~J013033.21+311447.0} is one of the longer period candidates. 
Although its folded light curve shows a well defined transit, its periodogram (Figure 10) 
has excess scatter and additional photometric data would be beneficial.
We derived a reasonable planetary radius of 1.07R$_J$ for its G4 spectral type.

{\bf 1SWASP~J015625.53+291432.5} The observed short period for this
system (1.45 days) may place it in the class of {\it very-hot-Jupiters}.
Its periodogram, showing a well defined period, and light curve are presented in Figure 11.
However, the transit depth is relatively shallow for a 0.7R$_J$ planet and
it has a large $\eta$, making it more likely to be a grazing incident
eclipsing stellar system or diluted by another star. There is another object
within the SW aperture making the later scenario more likely.
We note this as a marginal ESP candidate.

{\bf 1SWASP~J015711.29+303447.7} The light curve for this system is flat
with a well pronounced transit (Figure 12). We derived a planet radius of 1.37R$_J$ 
for its derived spectral type of F6. The expected transit duration 
is reasonable and this is a good ESP candidate worth further follow-up.

{\bf 1SWASP~J022651.05+373301.7} 
The short period of 1.2 days (Figure 13) may make this a {\it very-hot-Jupiter}, but 
its periodograms shows excess scatter and additional photometric measurements are
needed to improved the period determination. This source may also
be a low mass stellar system based on its large ellipsoidal amplitude (4.3).  
Its derived spectral type of F1 for this star results in a large planet of nearly 1.5R$_J$. 
It is one of the most interesting systems that will be follow-up spectroscopically.    

{\bf 1SWASP~J023445.65+251244.0} 
has a moderate amount of scatter in its light curve, but the transit
is well pronounced (Figures 14).  The transit depth and spectral types
result in similar sized planets of $\approx$1.2R$_J$. 
The expected transit duration is near the higher 
end for acceptable values ($\eta \approx$ 1.2), 
and further observations are needed to secure this source as ESP and possible
very-hot-Jupiter. 

{\bf 1SWASP~J025500.31+281134.0} shows a moderate amount of scatter in its
light curve (Figure~15), but does have a well defined period (2.2 days) and transit (3.7\%).
This transit depth is one of the larger for ESP candidates and results in a planet radius
of 1.3 R$_J$ for a spectral type of K0. The slight 'V' shape to the transit profile and large value of 
the expected transit duration ($\eta$ = 1.3) may indicate this is a stellar system. 

{\bf 1SWASP~J231533.56+232637.5}  and {\bf 1SWASP~J234318.41+295556.5} 
have a moderate amount of scatter in their light curves, but the transits
are well pronounced for both systems (Figures 16 \& 17).  Their transit depths and spectral types
result in similar sized planets of $\approx$1.1 and 1.4 R$_J$, respectively. 
1SWASP~J234318.41+295556.5 has an ellipsoidal amplitude of 2.9, and may 
be a stellar binary.

    \begin{figure} 
    \epsfig{file=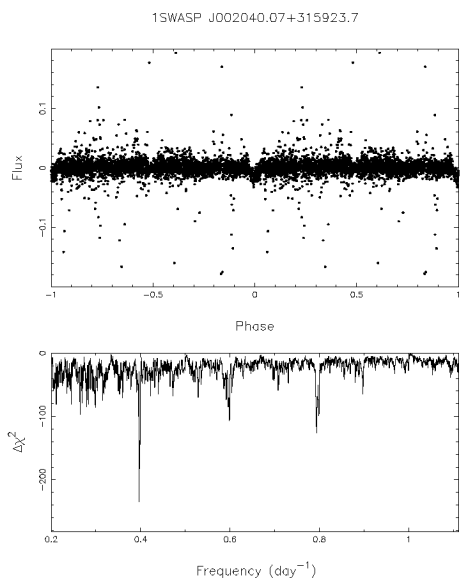, width=7cm}  
\caption{Timing data for the ESP candidate 
1SWASP~J002040.07+315923.7. 
The {\it top} panel shows the light curve folded on the best fit period,
and {\it bottom} panel shows the periodogram.
\label{fig6}} 
\end{figure}

 \begin{figure} 
    \epsfig{file=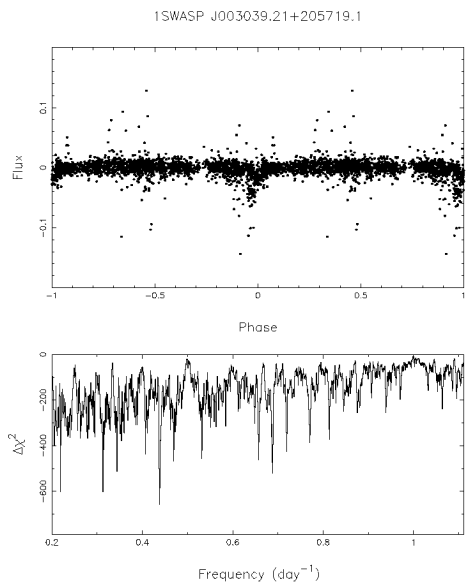, width=7cm}  
\caption{Timing data for the ESP candidate 
1SWASP~J003039.21+205719.1. Details are the same as Figure 6. 
\label{fig7}} 
\end{figure}
 \begin{figure} 
    \epsfig{file=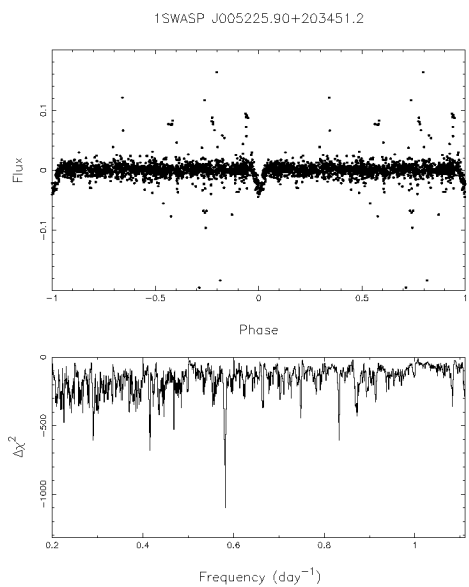, width=7cm}  
\caption{Timing data for the ESP candidate 1SWASP~J005225.90+203451.2. 
Details are the same as Figure 6. 
\label{fig8}} 
\end{figure}

 \begin{figure} 
    \epsfig{file=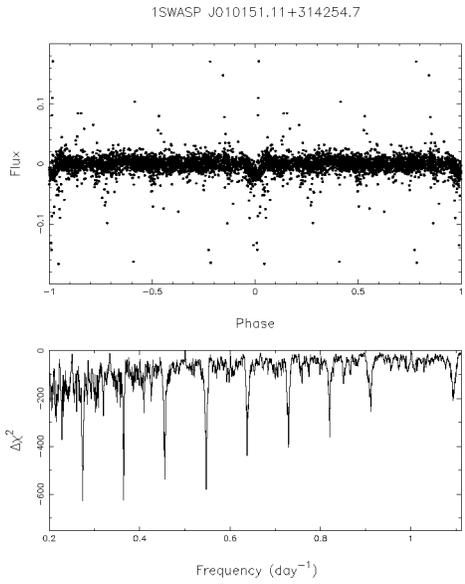, width=7cm}  
\caption{Timing data for the ESP candidate 
1SWASP~J010151.11+314254.7. Details are the same as Figure 6. 
\label{fig9}} 
\end{figure}

 \begin{figure} 
    \epsfig{file=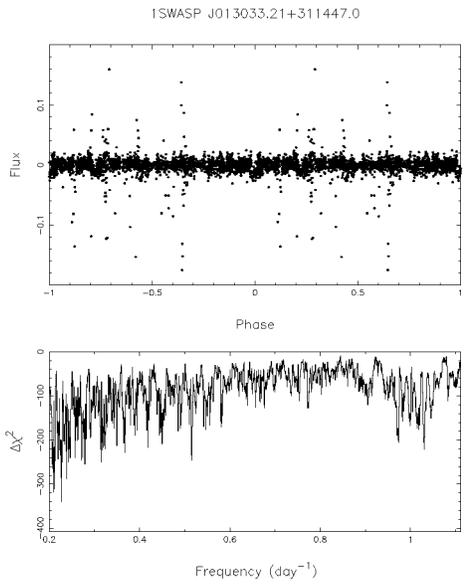, width=7cm}  
\caption{Timing data for the ESP candidate 
1SWASP~J013033.21+311447.0. Details are the same as Figure 6. 
\label{fig10}} 
\end{figure}

 \begin{figure} 
    \epsfig{file=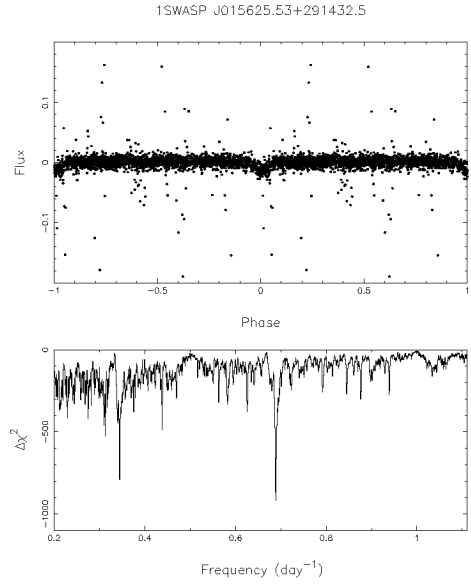, width=7cm}  
\caption{Timing data for the ESP candidate 
1SWASP~J015625.53+291432.5. Details are the same as Figure 6. 
\label{fig11}} 
\end{figure}

 \begin{figure} 
    \epsfig{file=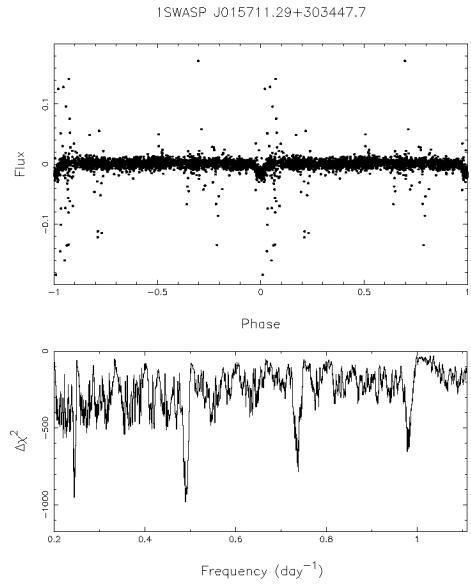, width=7cm}  
\caption{Timing data for the ESP candidate 
1SWASP~J015711.29+303447.7. Details are the same as Figure 6. 
\label{fig12}} 
\end{figure}

 \begin{figure} 
    \epsfig{file=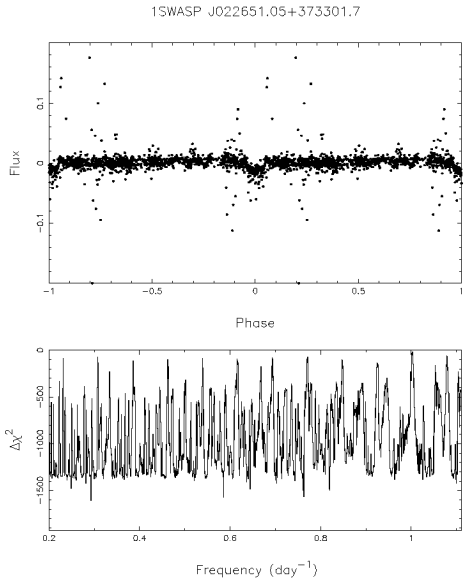, width=7cm}  
\caption{Timing data for the ESP candidate 
1SWASP~J022651.05+373301.7. Details are the same as Figure 6. 
\label{fig13}} 
\end{figure}

 \begin{figure} 
    \epsfig{file=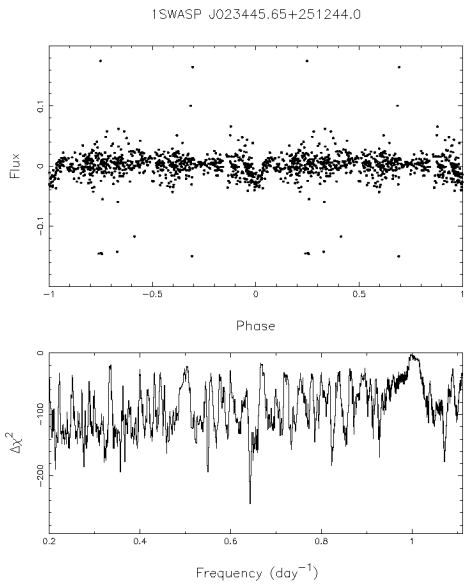, width=7cm}  
\caption{Timing data for the ESP candidate 
1SWASP~J023445.65+251244.0. Details are the same as Figure 6. 
\label{fig14}} 
\end{figure}

 \begin{figure} 
    \epsfig{file=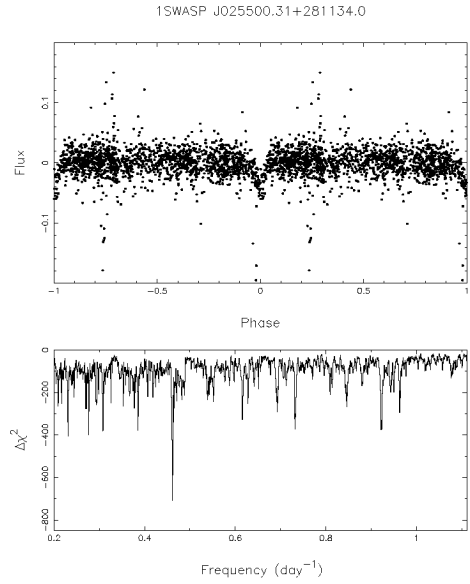, width=7cm}  
\caption{Timing data for the ESP candidate 
1SWASP~J025500.31+281134.0. Details are the same as Figure 6. 
\label{fig15}} 
\end{figure}

 \begin{figure} 
    \epsfig{file=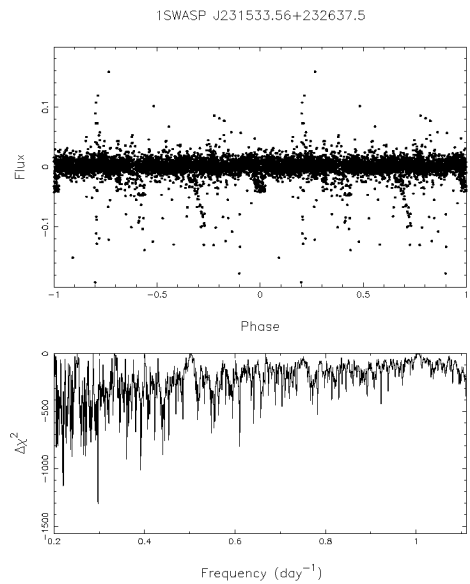, width=7cm}  
\caption{Timing data for the ESP candidate 
1SWASP~J231533.56+232637.5. Details are the same as Figure 6. 
\label{fig16}} 
\end{figure}

 \begin{figure} 
    \epsfig{file=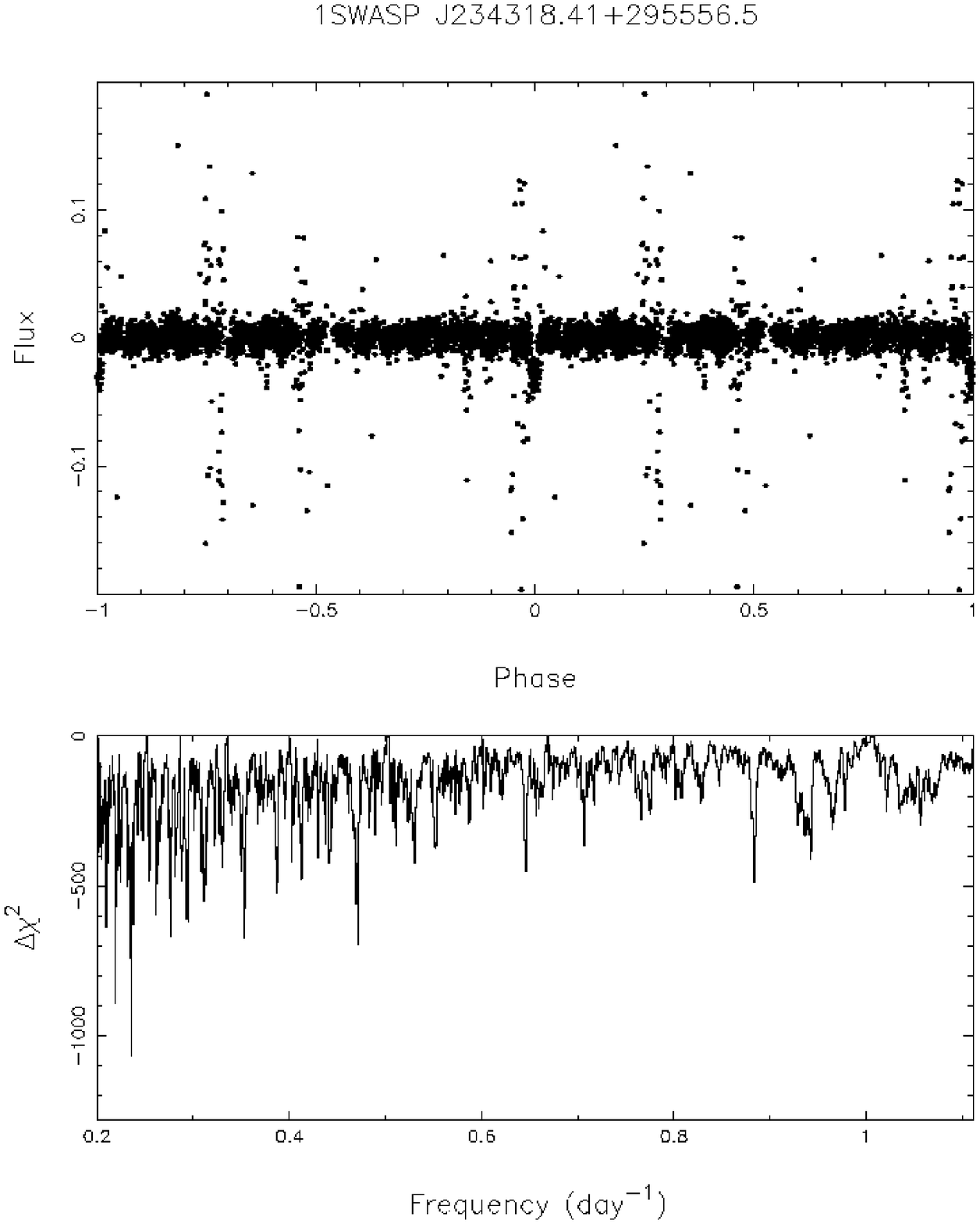, width=7cm}  
\caption{Timing data for the ESP candidate
1SWASP~J234318.41+295556.5. Details are the same as Figure 6. 
\label{fig17}} 
\end{figure}

\section{Discussion} 
The majority of candidates found with low amplitude periodic variability
showed obvious signatures of being stellar
binaries.  We have selected only 12 as 
new ESP candidates.
There are many different types of stellar systems and chance alignments 
that can show behavior mimicking a planet's transit.  Such possibilities
have been separated into 3 groups (see for example Brown 2003):
1) grazing incidence stellar binary systems, 2) stellar systems consisting
of both a high mass and low mass stars, and 3) an eclipsing stellar system
in which some light from a foreground or background star contaminates the
light from the binary reducing the depth of the eclipse and making it appear
more 'transit-like'.
We have scrutinized our sample of low amplitude variables with the
techniques outlined in \S\ 3.2, and attempted to eliminate
systems from above categories 1 and 2.
However, previous studies on ESPs have shown that it is difficult
to remove the 3rd category of false-positive transiting system, the 
diluted stellar binary \citep{To05, OD06}.
For this reason, high resolution optical spectroscopy is
needed to measure radial velocities for our candidates and 
determine the companion's mass. Such follow-up have been successful
in identifying $\approx$6 such sources from OGLE sample 
\citep{Kon03, Kon04, Bo04, Po04, Kon05}, and is being undertaken by our consortium.

\subsection{Expected Numbers of ESP}
For our sample of 900,000 stars we are left with 
$\approx$400,000 that are brighter than V=13.5 and for which
SuperWASP is statistically sensitive to detect a transiting ESP.
If we conservatively estimate only 50\% of these stars 
are late-type (F--M) dwarfs,
this reduces our sample size to $\sim$200,000 stars.  
If 1\% of these have hot-Jupiter companions \citep{LG03},  
and we expect 10\% of these to show transits \citep{H03}, 
we therefore should have detected $\approx$ 200 ESPs, 
although this estimate is affected by the true coverage
for a particular field. For this reason 
we have also estimated the expected fraction of transits recovered
for each observed field as a function of transit period and for different
numbers of transits. These expected transit yields are shown in 
Figure~18a-c for the recovery of 2, 4, and 6 transits. 
For sources with periods $\leq$5 days and
for fields that were observed for 80 or more nights, 2 transits would
have been detected $\approx$95\% of the time. For the fields with poorest coverage
the recovered fraction only drops to $\approx$80\%. 
However, if we require the recovery of 4 transits, the recovered fraction
is reduced to $\approx$40\% for fields that were observed for at least 80 nights,
and $\approx$10--20\% for fields with $\leq$60 nights of coverage. 
The majority of the fields
presented here were observed for at least 60 nights, and 
hence we should recover 2 transits for $\approx$80\% of the planets present, leading to
an expectation of 160 ESPs detected. If we require the recovery of 4 transits,
then this number is reduced to $\approx$80. 

 \begin{figure} 
    \epsfig{file=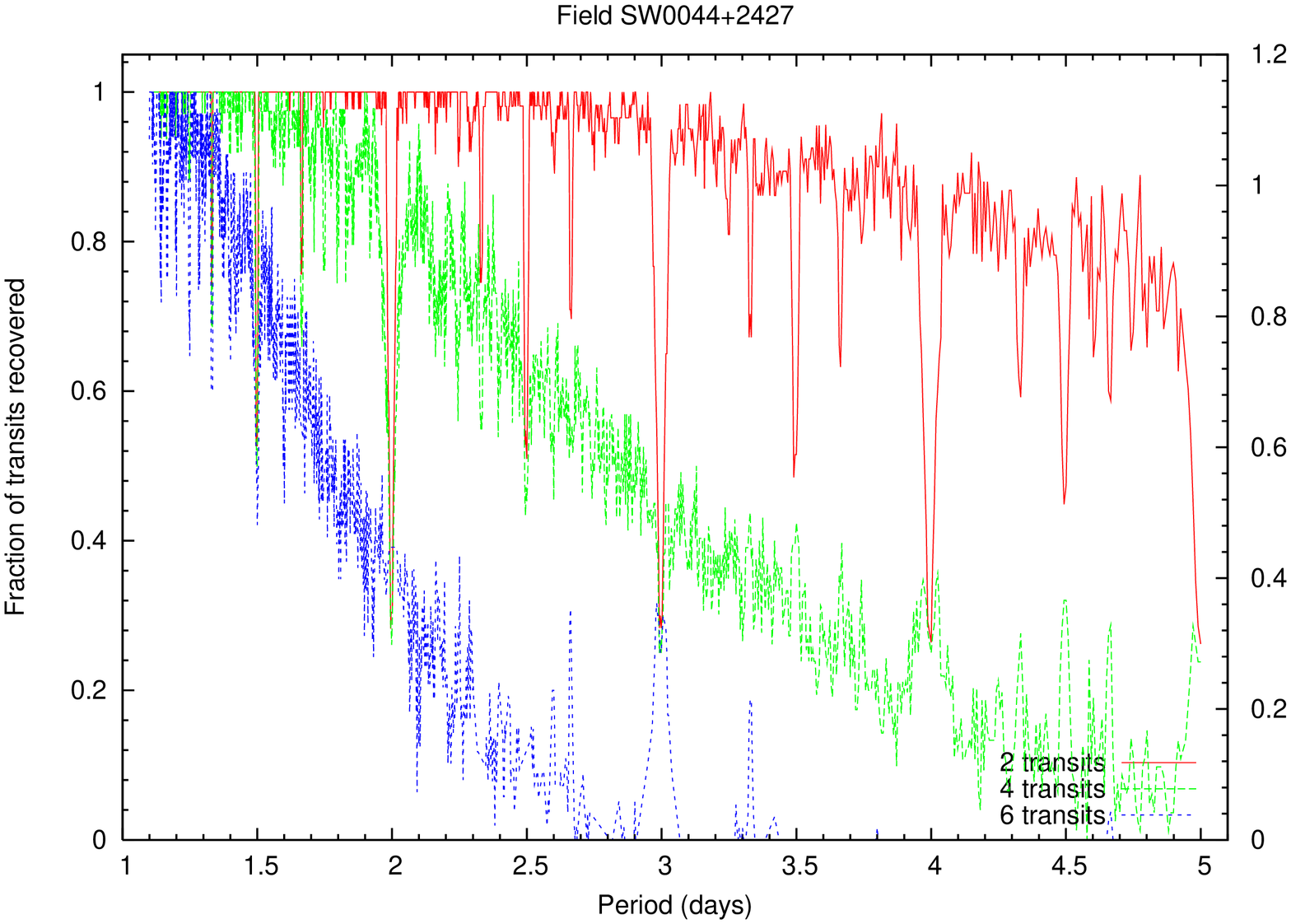, width=6cm, angle=-90} 
    \epsfig{file=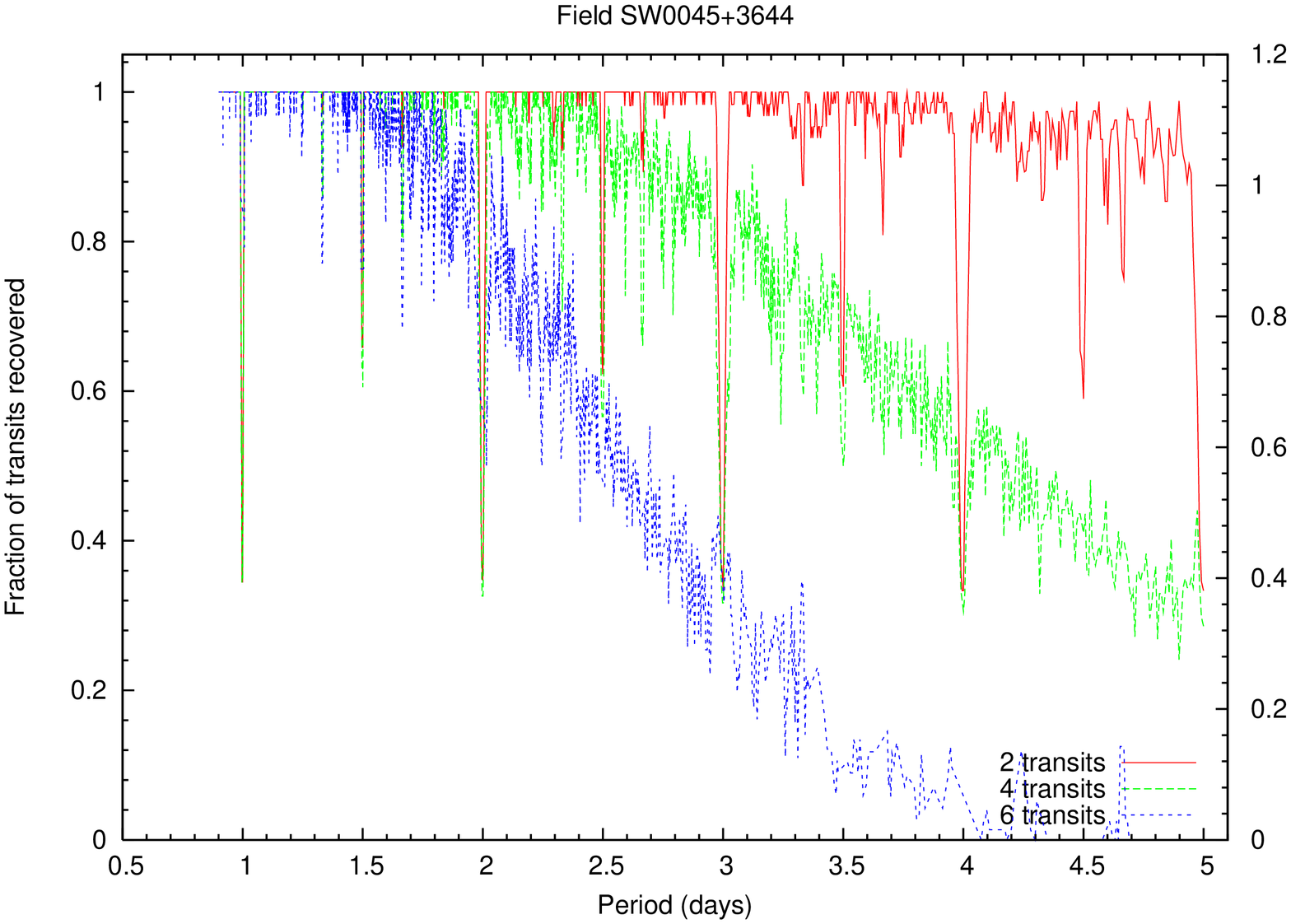, width=6cm, angle=-90} 
    \epsfig{file=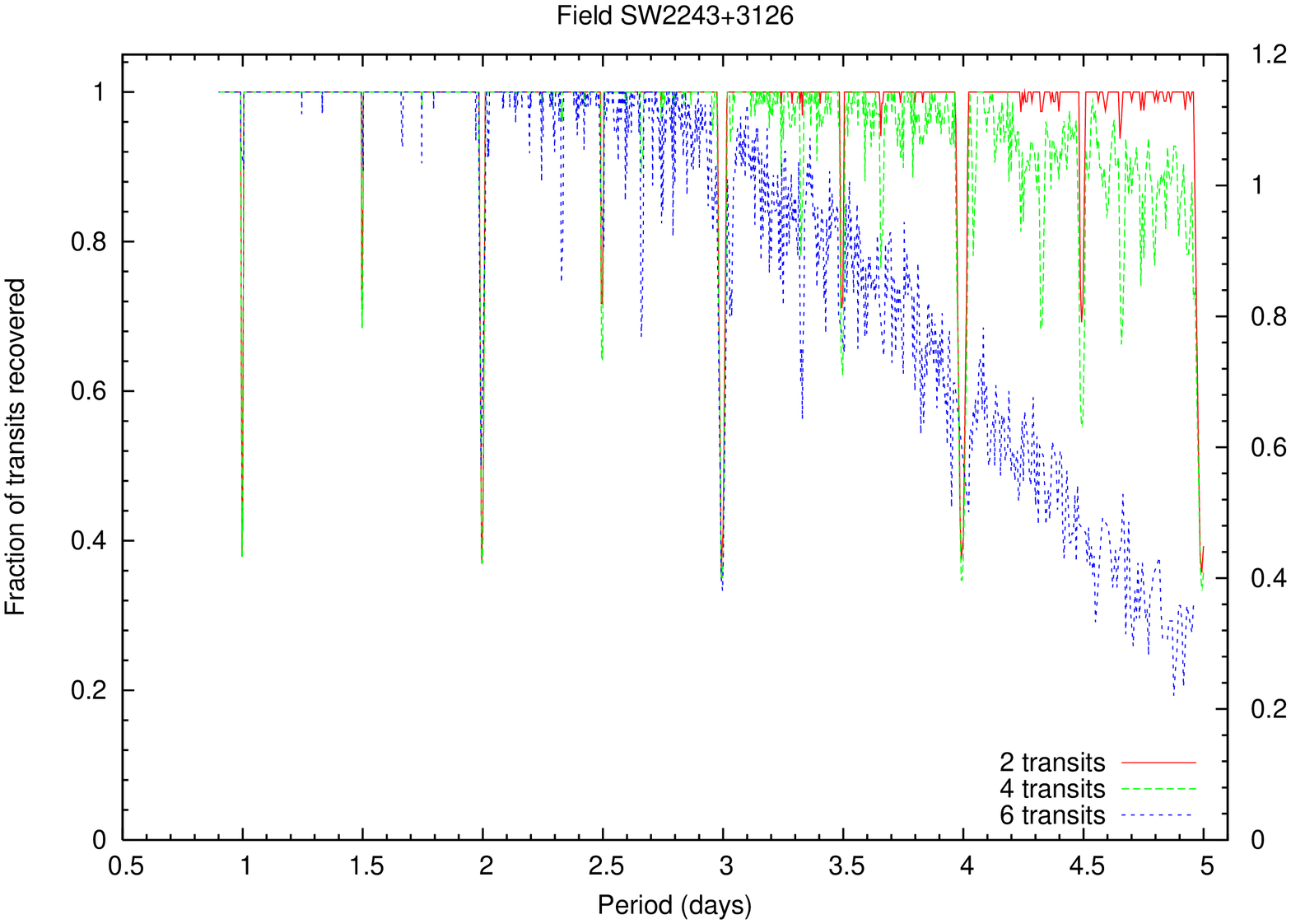, width=6cm, angle=-90}  
\caption{ The expected fraction of transits recovered as a function
of period and for the detection of 2, 4, and 6 transits.
We show the expected fraction recovered for three fields: a. 
0044+2427, b. 0045+3644, and c. 2243+3126, representing coverages of 
$\approx$30, 80, and over 100 nights, respectively.
\label{fig18}} 
\end{figure}

 The plethora of short-period extra-solar planets that was predicted
in the mid-1990s and recently \citep{H03} has not come to fruition.
Surveys that concentrate on wide fields and brighter stars
have only 2 transit ESP discoveries (TrES-1, Alonso et al. 2004, and XO-1b, McCullough et al. 2006). 
The majority of other 
transiting ESPs (6) have come from the OGLE micro-lensing survey of the 
Galactic Centre and fields nearby \citep{U04}. \citet{Br03} estimated the detection 
rate for
giant ESP including the probability both of a star having a planet and 
of observing it. This provides an estimate of $\approx$1 ESP
per 25,000 stars.  Photometric surveys of open clusters (OCs) with 2 to 4 metre class
telescopes that observe to much fainter magnitudes 
[e.g. \cite{S03}; EXPLORE, \cite{vB05}; PISCES, \cite{Moch05, Moch06}] 
have also failed to detect short-period ESPs. 
These surveys have monitored tens of thousands of stars
for tens of days with high precision and to very faint magnitudes (R $\sim$ 20),
yet found few to no ESPs. 

A trend has been established where ESP are found predominately with 
metal-rich host stars \citep{FV05, SI03}. The lack of ESP transit candidate detections
in metal-poor globular cluster, 47~Tuc \citep{G00, W05}, 
is consistent with this.
Given that these deep OC surveys are sensitive to short period ESPs, the 
lack of detections might imply that the frequency of transiting hot-Jupiters is closer
to 1 in 50,000 stars.  
Our present results of 12 ESP candidates in nearly 400,000 stars are consistent
with the current detection rates.

There may be several contributing factors to this lack of detections.
The weather can seriously diminish the observing duration and shorten
the observing window used in many of the estimated ESP yields.
Changes in extinction can be difficult to remove from the
observations and cause false variability. Additionally, intrinsic stellar 
variability, especially in stars of later spectral types, 
may hide the transit signature.
Later type stars will have deeper transit depths 
and these are more easily recovered for fainter stars.
For these reasons, we may then consider if the periodic nature of the transit signal would
allow us to recover shallower transit depths or transits from stars
fainter than $\approx$13th magnitude. 
Periodic signals are most readily recovered from the transit searching
algorithm, but we are ultimately limited by the signal-to-noise and
duration of the data. 
Simulations for the fraction of transits recovered based on each of our field's
observing windows have shown that there are several integer periods where the
probability of recovering a transit is greatly reduced because of the lack of coverage
(e.g. Figure 18a-c), and there also several periods where the chance for
detection is actually increased (see Fig 18a, N$_{tr}$ = 6 case and P=3 days).
In a companion paper, \citet{Sm06} investigate the affects of the noise
properties on transit recovery in the SuperWASP data and conclude
the best way to reduce the co-variant noise is to increase the observing baseline.

Many improvement to the analysis of photometric transit data have
been made, and there are additional tests to distinguish stellar systems
from true ESPs (e.g. Tingley \& Sackett 2005). 
The ESP detection rate of the WASP project is presently unknown, but 
continued SW-N and SW-S observations will discover more high quality ESP candidates.  
However, it is important
to have spectroscopic follow-up to determine the mass function of these
systems.  
It may be left to space-borne missions, such as Kepler and COROT,
without the hindrance of the atmosphere, to truly determine the
frequency of hot-Jupiter and even Earth-size ESPs.

\begin{table*} 
 \centering
 \begin{minipage}{180mm}
\vspace{-0.1in}
\caption{Light Curve data for SuperWASP Low Amplitude Sources} 
\begin{tabular}{lcccccccccr} 
\hline 
\hline 
 Name & S$_{red}^a$ & Period & $\delta$ & duration &  Epoch$^b$ & N$_{trans}$ & $\Delta\chi^{2}$ & (S/N)$_{ellip}^c$   \\
 1SWASP  & & (days)   & (percent)  & (hrs) &  &  &   \\\hline 
  J230808.34+333803.9 & 21.2 &2.23 &2.15 &2.45 &3149.2152& 15&  1922& 7.3   \\
  J231240.14+321540.4 & 12.6 & 2.53 & 3.08 & 2.59 & 3148.7063 & 9 & 969 & 2.10 \\ 
  J231302.08+262724.3 & 11.1 &4.39 &2.16 &3.05 &3149.1804&  7&  1188& 4.0  \\
  J231533.56+232637.5 & 11.6 &3.36 &2.89 &3.79 &3148.7560& 10&  1597& 0.7   \\
  J231807.73+240522.1 & 21.3 & 2.54 & 5.96 & 1.90 & 3150.5115 & 8 & 1281 & 1.5 \\
  J232639.10+233219.2 & 16.8 &2.21 &3.67 &2.54 &3150.3117& 15&  3405& 2.3  \\
  J232700.56+200609.1 & 22.3 &3.24 &5.79 &2.45 &3149.3855& 11&   733& 3.2  \\
  J233325.17+332632.1 & 12.4 &2.27 &1.59 &3.34 &3150.2111& 16&  1072& 9.2   \\
  J234318.41+295556.5 & 14.4 &4.24 &2.81 &2.42 &3151.8626&  6&  1213& 2.9  \\
  J000029.25+323300.9 &14.0 &1.20 &2.49 &2.35 &3151.9736& 17&   939& 7.2  \\
  J000233.27+331516.8 &15.6 &2.37 &3.01 &2.04 &3152.5501& 11&  1476& 10.2   \\
  J002040.07+315923.7 &11.6 &2.52 &1.27 &3.00 &3151.4864&  9&   287&  0.4   \\
  J002728.02+284649.2 &9.7 &1.52 &1.22 &3.17 &3166.9666& 12&  1136&  2.5  \\
  J002819.22+335122.1 &12.0 &3.45 &8.07 &3.82 &3165.2148&  6&  1033&  2.6   \\
  J003039.21+205719.1 & 10.5 &2.28 &1.91 &2.38 &3166.6460&  9&   668& 0.4  \\
  J004804.21+202258.8 & 25.2 &1.83 &6.45 &2.71 &3167.0060&  9&  3326&  6.6  \\
  J005107.84+214352.7 & 13.7 &1.47 &4.24 &2.42 &3167.6596&  8&  3322&  5.9   \\
  J005225.90+203451.2 & 12.4 &1.72 &2.78 &1.66 &3166.9509&  8&  1198&  3.7  \\
  J005250.45+221038.8 & 19.7 &2.46 &6.46 &2.30 &3167.0449&  6&  1011& 1.6   \\
  J005818.26+204348.0 & 10.8 &2.20 &8.83 &2.54 &3167.1818&  7&  2962& 7.0   \\
  J010151.11+314254.7 & 9.6 &2.19 &1.66 &2.64 &3167.4457& 10&   694& 3.2    \\
  J010553.38+241358.3 & 8.9&3.35 &2.88 &1.73 &3165.6700&  4&   305&  1.7   \\
  J010706.32+313918.0 & 20.2 & 4.10 & 4.54 & 2.14 & 3167.5442 & 4 & 2569 & 0.2 \\ 
  J013033.21+311447.0 & 7.0 &4.40 &1.56 &3.58 &3179.5786&  3&  408& 2.3   \\
  J013100.45+374745.2 & 15.6 &1.65 & 3.60 & 2.14 & 3181.5542 & 12 &  841 & 0.1  \\ 
  J013250.08+194332.7 & 10.7 &3.13 &5.60 &4.27 &3166.2451&  6&  1248 & 2.8 \\
  J013452.76+293626.5 & 11.8 &1.91 &1.53 &3.31 &3166.0957& 14&  1674& 1.1  \\
  J014400.22+344449.2 & 11.9  &3.72 &2.73 &3.17 &3180.2839&  5&   914& 4.4   \\
  J015625.53+291432.5 & 12.2 &1.45 &1.24 &3.22 &3182.5415& 13&  1084& 4.9  \\
  J015711.29+303447.7 & 12.3 &2.04 &1.55 &2.30 &3182.5612&  9&  1253& 1.3   \\
  J015951.59+354455.4 & 8.5 &1.19 &2.28 &1.97 &3181.9274& 13&   586&  12.9   \\
  J020720.96+325526.5 & 10.1 &1.54 &4.52 &3.22 &3181.7719& 10&  1354&  5.8  \\
  J021217.50+335319.2 & 20.0 &3.91 &6.85 &2.64 &3180.5364&  3&  1883&  1.0  \\
  J022421.03+375419.9 & 11.3 &4.16 &5.47 &2.81 &3191.8221&  4&  2530&  3.3  \\
  J022651.05+373301.7 & 9.6 &1.22 &1.29 &2.02 &3194.6500&  8&  1782&  4.3  \\
  J023445.65+251244.0 & 7.6 &1.55 &2.36 &2.62 &3194.4315&  6&   286&  1.1  \\
  J024206.53+364029.7 & 16.4 &2.59 &4.78 &2.81 &3193.8696&  7&   763& 3.4   \\
  J025419.14+324240.7 & 20.4 &2.91 &5.11 &2.71 &3192.4337&  4&   319& 0.4  \\
  J025500.31+281134.0 & 9.0 &2.17 &3.74 &2.76 &3191.9254&  6&   767& 1.4   \\
  J025712.69+403140.0 & 9.8 &2.22 &1.80 &2.50 &3192.7651&  6&   549& 10.7  \\
  J025958.91+294434.6 & 10.0 &1.73 &5.41 &2.33 &3193.1976&  6&  1010& 2.3  \\
\hline
\end{tabular}
\\
$^a$ S$_{red}$ -- Signal-to-red noise (see text). \\
$^{b}$ JD = 2450005.5 + Epoch \\
$^c$ (S/N)$_{ellip}$ -- Amplitude of ellipsoidal variations (see text). \\
\end{minipage}
\end{table*}

\section{Conclusions}
We have presented 41 low amplitude variables 
from the SW-N 2004 observing season. From this list we identified 
12 ESPs candidates with periods of 1.2 to 4.4 days. These 
have transit depths $\approx$1--4\% implying planetary radii of 1.0--1.5 R$_J$ 
for their derived spectral
types. High resolution optical spectroscopy to measure radial 
velocities and obtain orbital solutions and masses for these systems are needed
to confirm these candidates are true extrasolar planets.  
Such observations are being undertaken by our consortium.
SW-N and SW-S both continue to operate for the 2006 observing season, and
with a full 8 cameras complement, and are expected to obtain more than 
twice the amount of data as the 2004 campaign.

\begin{table*} 
 \centering
 \begin{minipage}{140mm}
\caption{SuperWASP Candidate Stellar and Planetary Parameters} 
\begin{tabular}{lcccccccccr} 
\hline 
\hline 
  Name & V$_{SW}$ &  V$-$K& J$-$K & T$_{eff}$ & Spec. Type  & R$_{\star}$ & R$_{p}$ &  $\eta$ & Comment    \\
 1SWASP&          &       &       &           &             & R$_{\sun}$ & R$_J$  &   &   \\\hline

 J230808.34+333803.9 & 10.8 &  1.41 & 0.28 &  6051 & F9 & 1.14 & 1.43 & 0.85 & Ellip. large \\ 
 J231240.14+321540.4 & 11.8 & 1.46 & 0.30 & 5990 & G0 & 1.11 & 1.66 & 0.85 & R$_{p}$ too large \\ 
 J231302.08+262724.3 & 10.6 &  3.68 & 0.87 &  4170 & M0 & 0.63 & 0.79 & 1.19 & Giant? \\ 
 {\bf J231533.56+232637.5} & 11.6 &  1.72 & 0.35 &  5661 & G6 & 0.96 & 1.39 & 0.13 & ESPc \\ 
 J231807.73+240522.1 & 12.0 & 2.31 & 0.54 & 5034 & K3 & 0.76 & 1.63 & 0.74 & R$_{p}$ too large \\
 J232639.10+233219.2 & 11.5 &  1.26 & 0.22 &  6256 & F7 & 1.25 & 2.04 & 0.81 & R$_{p}$ too large \\ 
 J232700.56+200609.1 & 12.5 &  1.55 & 0.29 &  5870 & G2 & 1.05 & 2.16 & 0.73 & R$_{p}$ too large \\ 
 J233325.17+332632.1 & 11.1 &  1.74 & 0.41 &  5637 & G6 & 0.95 & 1.01 & 1.25 & Ellip. large \\ 
{\bf  J234318.41+295556.5} & 10.7 &  2.31 & 0.55 &  5034 & K3 & 0.76 & 1.09 & 0.84 & ESPc \\ 
 J000029.25+323300.9 & 12.0 &  2.24 & 0.57 &  5101 & K2 & 0.77 & 1.01 & 1.29 & Ellip. large \\ 
 J000233.27+331516.8 & 11.3 &  1.04 & 0.29 &  6573 & F4 & 1.40 & 2.07 & 0.60 & R$_{p}$ too large \\ 
 {\bf J002040.07+315923.7} & 11.8 &  1.51 & 0.31 &  5921 & G1 & 1.08 & 1.04 & 1.06 & ESPc \\ 
 J002728.02+284649.2 &  9.6 &  2.84 & 0.71 &  4602 & K5 & 0.69 & 0.65 & 1.72 & Giant?/Large $\eta$ \\ 
 J002819.22+335122.1 & 12.5 &  1.52 & 0.35 &  5908 & G1 & 1.07 & 2.39 & 0.98  & R$_{p}$ too large\\ 
 {\bf J003039.21+205719.1} & 11.0 &  1.31 & 0.55  &  6186 & F8 & 1.21 & 1.23 & 0.92 & ESPc \\ 
 J004804.21+202258.8 & 10.7 &  2.08 & 0.50 &  5260 & K1 & 0.82 & 1.78 & 1.1 & Ellip. large \\ 
 J005107.84+214352.7 & 11.1 &  1.51 & 0.30 &  5921 & G1 & 1.08 & 1.90 & 0.95  & R$_{p}$ too large\\ 
 {\bf J005225.90+203451.2} & 11.2 &  1.89 & 0.46 &  5465 & G9 & 0.88 & 1.25 & 0.71 & ESPc \\ 
 J005250.45+221038.8 & 12.3 &  1.47 & 0.33 &  5973 & G0 & 1.10 & 2.38 & 0.74 & R$_{p}$ too large \\ 
 J005818.26+204348.0 & 12.4 &  1.77 & 0.38 &  5602 & G7 & 0.93 & 2.09 & 0.63 & R$_{p}$ too large \\ 
 {\bf J010151.11+314254.7} & 11.1 &  1.04 & 0.17 &  6573 & F4 & 1.40 & 1.42 & 1.04 & ESPc \\ 
 J010553.38+241358.3 & 11.1 &  1.49 & 0.33 &  5947 & G1 & 1.09 & 1.58 & 0.53 & R$_{p}$ large \\ 
 J010706.32+313918.0 & 10.3 & 1.55 & 0.30 & 5870 & G2 & 1.05 & 1.91 & 0.6 & R$_{p}$ too large \\ 
 {\bf J013033.21+311447.0} & 10.6 &  1.64 & 0.34 &  5758 & G4 & 1.00 & 1.07 & 1.08 & ESPc \\ 
 J013100.45+374745.2 & 11.1 & 1.35  & 0.30 & 6130 & F8 & 1.18 & 1.90 & 0.78 & R$_{p}$ large  \\ 
 J013250.08+194332.7 & 11.4 &  1.91 & 0.42 &  5442 & G9 & 0.87 & 1.76 & 0.14 & R$_{p}$ too large \\ 
 J013452.76+293626.5 & 10.1 &  2.68 & 0.64 &  4720 & K5 & 0.70 & 0.40 & 1.60 & Giant/Large $\eta$ \\ 
 J014400.22+344449.2 & 11.2 &  1.30 & 0.25 &  6200 & F8 & 1.22 & 1.72 & 0.88 & R$_{p}$ too large \\ 
 {\bf J015625.53+291432.5} & 10.3 &  2.30 & 0.54 &  5044 & K3 & 0.76 & 0.72 & 1.67 & ESPc?/Large $\eta$ \\ 
 {\bf J015711.29+303447.7} & 10.4 &  1.19 & 0.34 &  6354 & F6 & 1.30 & 1.38 & 0.77 & ESPc \\ 
 J015951.59+354455.4 & 11.0 &  3.07 & 0.73 &  4453 & K7 & 0.67 & 0.86 & 0.08 & Ellip. large \\ 
 J020720.96+325526.5 & 12.4 &  2.27 & 0.65 &  5072 & K2 & 0.77 & 1.40 & 1.51 & Ellip. large \\ 
 J021217.50+335319.2 & 11.2 &  2.62 & 0.67 &  4767 & K5 & 0.71 & 1.59 & 0.92 & R$_{p}$ large  \\ 
 J022421.03+375419.9 & 10.7 &  1.70 & 0.30 &  5685 & G5 & 0.97 & 1.94 & 0.81 & R$_{p}$ too large \\ 
 {\bf J022651.05+373301.7} &  8.3 &  0.83 &0.11 &  6896 & F1 & 1.52 & 1.47 & 0.74 & ESPc \\ 
 {\bf J023445.65+251244.0} & 12.1 &  1.86 & 0.35 &  5498 & G8 & 0.89 & 1.17 & 1.17 & ESPc \\ 
 J024206.53+364029.7 & 11.8 &  1.47 & 0.29 &  5973 & G0 & 1.10 & 2.05 & 0.89 & R$_{p}$ too large \\ 
 J025419.14+324240.7 & 11.7 &  1.61 & 0.37 &  5795 & G3 & 1.02 & 1.97 & 0.86 & R$_{p}$ too large \\ 
 {\bf J025500.31+281134.0} & 11.8 &  2.00 & 0.45 &  5344 & K0 & 0.84 & 1.34 & 1.29 & ESPc \\ 
 J025712.69+403140.0 & 10.6 &  1.13 & 0.17 &  6441 & F6 & 1.34 & 1.53 & 0.80 & Ellip. large \\ 
 J025958.91+294434.6 & 12.2 &  1.57 & 0.21 &  5845 & G2 & 1.04 & 2.11 & 0.71 & R$_{p}$ too large \\ 

\hline
\end{tabular}
\end{minipage}
\end{table*}

\section*{Acknowledgments}
The WASP consortium consists of representatives from the Queen's
University Belfast, University of Cambridge (Wide Field Astronomy Unit), 
Instituto de Astrof\'isica de Canarias,  Isaac Newton Group of Telescopes (La Palma),  University of
Keele, University of Leicester, The Open University, and the University of St Andrews. 
The SuperWASP-N and SuperWASP-S instruments were constructed and operated 
with funds made available from Consortium Universities   
and the Particle Physics and Astronomy Research Council. 
SuperWASP North is located in the Spanish Roque de Los Muchachos Observatory on La Palma,  
Canary Islands which is operated by the Instituto de Astrof\'isica de Canarias (IAC). 
We thank an anonymous referee for suggested improvements, and
acknowledge useful scientific discussion with I. Dino, J. Smoker, and C. Snodgrass.
We are grateful to PPARC for financial support. 



\end{document}